\documentstyle[12pt,epsfig,epsf]{article}
\newcommand{\dst}{\displaystyle}

\newcommand{\institute}[1]{\parbox{16cm}{%
\centering\normalsize \sl #1}}
\vspace{2cm}

\title{
\bf Static Potential in the SU(2)-Higgs Model and Coupling Constant 
Definitions in Lattice and Continuum Models  }
\author{%
F.~Csikor, Z.~Fodor, P.~Heged\"us, A.~Pir\'oth \\
\institute{Institute for Theoretical Physics, E\"otv\"os University,\\
H-1088 Budapest, Hungary }\\
}
\date{}
\begin{document}
\maketitle

\vspace{3.cm}
\begin{abstract}
We present a one-loop calculation of the static potential in the SU(2)-Higgs 
model. The connection to the coupling constant definition used in 
lattice simulations is clarified. The consequences in comparing lattice 
simulations and perturbative results for finite temperature applications 
are explored.

\end{abstract}
%\vfill

\section{Introduction}

The observed baryon asymmetry of the universe was eventually determined
at the electroweak phase transition \cite{KRS85}. The most straightforward
method to study this phase transition is to use resummed
perturbation theory (cf.\ e.g.\ \cite{AE93,BFHW94,FH94}).
In the low temperature Higgs phase
the perturbative approach is expected to work well, however, it is
not able to describe the high temperature symmetric phase, which has serious
infrared problems in perturbation theory. Since the determination of
thermodynamical quantities at the critical temperatures is based on the
properties of both phases, non-perturbative techniques are necessary for a
quantitative understanding of the phase transition.

One very succesful
possibility is construct an effective 3-dimensional theory by using
dimensional reduction, which is a perturbative step. The non-perturbative study
is carried out in this effective 3-dimensional model (see e.g.\ \cite{3d-sim}
and references therein). Analytical estimates are
confirmed by numerical results and relative errors are believed to be at the
percent level.

Another approach is to use 4-dimensional simulations. The complete lattice 
analysis of the Standard Model is not feasible due to the presence of chiral
fermions, however, the infrared problems are connected only with the bosonic
sector. These are the reasons why the problem is usually studied by simulating
the SU(2)-Higgs model on 4-dimensional lattices, and perturbative steps are used
to include the U(1) gauge group and the fermions. Finite temperature simulations
are carried out on lattices with volumes $L_t \cdot L_s^3$, where $L_t \ll L_s$
are the temporal and spatial extensions of the lattice, respectively.
The lattice spacing is basically fixed by the number of the lattice
points in the temporal direction
($T_c=1/(L_t a)$, where $T_c$ is the critical temperature in physical units);
therefore huge lattices are needed to study the soft modes. This
problem is particularly severe for Higgs boson masses around the W mass,
for which the phase transition is weak and typical correlation lengths
are much larger than the lattice spacing. In this case asymmetric lattice
spacings are used, in particular the spatial lattice unit is approximately 
four times larger than the temporal one \cite{4d-asym}.

Despite the fact that the two approaches (perturbative and lattice)
are systematic and well-defined, it is not easy to compare
their predictions. The reason for this is that in lattice simulations
the gauge coupling constant is determined from the static potential, whereas
in perturbation theory the ${\overline {\rm {MS}}}$ scheme is used. The main goal of this
paper is to perform as perfect a comparison as possible, by determining the
${\overline {\rm {MS}}}$ gauge coupling constants,
which correspond to the different lattice results.\footnote{
During the write-up of our results, prior to us, a similar,
independent calculation for the gauge coupling constant was
presented by M.~Laine \cite{Laine}, who compared 4-dimensional and 3-dimensional results, too. 
Using his convention for the renormalized gauge coupling, which is a special 
case of our definition, the two results agree. However, as it will be
discussed later, our definition for the perturbative gauge couplings is 
%slightly 
closer to the actual lattice definitions.}

The paper is organized as follows. Section 2 contains the one-loop
static potential of the SU(2)-Higgs model using the ${\overline {\rm {MS}}}$ scheme
in the Feynman gauge. Section 3 relates the continuum version of the
lattice gauge coupling constant to the ${\overline {\rm {MS}}}$ coupling. In Section 4
the detailed comparison of lattice and perturbative predictions are
presented. Section 5 summarizes our results.

\section{Calculation of the one-loop static potential}

The one-loop static potential was calculated long ago in quantum chromodynamics 
\cite{Suss,Fish,Appel}, and even the full two-loop result was published recently
\cite{Markus}. The calculation is based on the same principles and techniques in the
case of the SU(2)-Higgs model. 
One calculates rectangular Wilson loops of size $r\times t$. The logarithm 
divided by $-t$ gives the potential at distance $r$ in the $t \rightarrow 
\infty$ limit. 

Our calculation was performed in the ${\overline {\rm {MS}}}$ scheme and  
the Feynman gauge but the result is gauge independent, as it should be for a
physical observable.
The relevant graphs are shown in Fig.\ 1.
Other graphs, giving vanishing contributions in the Feynman gauge and are not
shown in Fig.\ 1.
Solid lines represent the heavy quark (antiquark) propagator, while 
wavy lines the vector boson propagator. External heavy quark 
(antiquark) propagators are not shown in the figure. The one-loop corrected 
vector boson propagator contains scalar and ghost contributions as well.
The result can be conveniently given in momentum space. One obtains

\begin{eqnarray}\label{mom_pot}
&\dst{V_{{\mathrm{1-loop}}}(k)=- \frac{3g^4}{32\pi^2}\frac{1}{k^2+M_W^2}} \nonumber \\
&\dst{\left\{\frac{k^2+M_W^2}{k}\frac{2}{\sqrt{k^2+4M_W^2}}
\log \frac{\sqrt{k^2+4M_W^2}-k}{\sqrt{k^2+4M_W^2}+k} \right.} \nonumber \\
&\dst{+\frac{1}{k^2+M_W^2}\left[\frac{1}{24R_{HW}^2}\left(86R_{HW}^2 k^2 
-9(6-3 R_{HW}^2 +R_{HW}^4 )M_W^2\right)\log\frac{\mu^2}{M_W^2} \right.} \nonumber \\
&\dst{+ \frac{1}{8} (13 k^2-20 M_W^2 ) F(k^2;M_W^2,M_W^2)}  \nonumber \\
&\dst{-\frac{1}{24}\left( (R_{HW}^2-1)^2 \frac{M_W^4}{k^2} +k^2 +2(R_{HW}^2-5)
M_W^2\right) F(k^2;M_W^2,M_H^2)} \nonumber \\
&\dst{+\frac{R_{HW}^2 \cdot \log R_{HW}}{12(R_{HW}^2-1)} \left( k^2+(9R_{HW}^2-17)M_W^2\right)} \nonumber \\
&\dst{\left. \left. +\frac{1}{72R_{HW}^2}\left( R_{HW}^2 k^2+3 (-18+R_{HW}^2-11 R_{HW}^4)M_W^2\right) \right] 
\right\} },
\end{eqnarray}
where $k^2$ denotes the square of the three-momentum $\vec{k}$, $M_H$ the 
Higgs mass and  $R_{HW}=M_H /M_W $. The function $F$ is defined as

\begin{eqnarray}\label{fv}
F(k^2;m_1^2,m_2^2)=1+\frac{m_1^2 +m_2^2 }{m_1^2 -m_2^2 } \log \frac{m_1}{m_2} +
\frac{m_1^2 -m_2^2 }{k^2} \log \frac{m_1}{m_2} \nonumber \\
+\frac{1}{k^2} \sqrt{ (m_1 +m_2 )^2 +k^2 )((m_1 -m_2 )^2 +k^2 )} \log 
\dst{\frac{1-\sqrt{\frac{(m_1 -m_2 )^2 +k^2}{(m_1 +m_2 )^2 +k^2 }}}
{1+\sqrt{\frac{(m_1 -m_2 )^2 +k^2}{(m_1 +m_2 )^2 +k^2 }}}}.
\end{eqnarray}
As it can be seen, our result does depend on the renormalization scale 
$\mu$ and it fully agrees with that of M.~Laine \cite{Laine}.

Eq.\ (\ref{mom_pot}) has to be Fourier transformed into coordinate space. 
We applied the brute force method performing numerical integration. As 
a check, we compared our results with various pieces of the partly analytic 
calculation in \cite{Laine} for the derivative of the potential 
(with respect to distance). The agreement is excellent.

\begin{figure}
\begin{center}
\epsfig{file=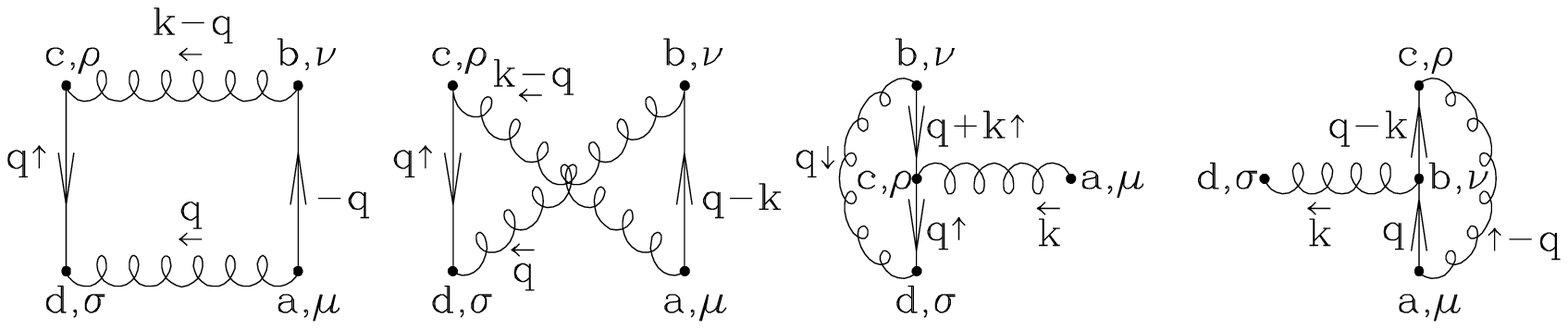,width=10.0cm}
\epsfig{file=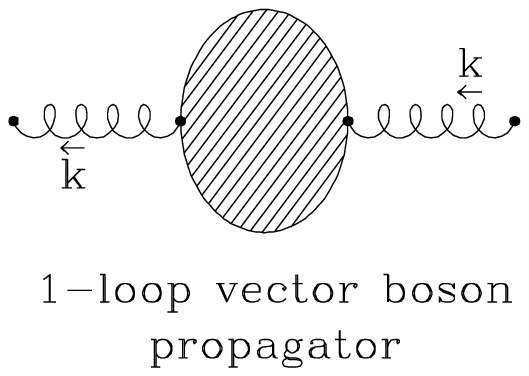,width=3.16cm}
\caption{\label{fig1}
{Graphs giving nonvanishing contributions to the static potential 
}}
\end{center}\end{figure}
Our result is presented in Figs.\ 2 and 3, where the various parts of the 
one-loop correction to the potential are  plotted. We define

\begin{eqnarray}\label{pot}
\frac{V(r)}{M_W}=-\frac{3g^2}{16\pi} \frac{\exp(-M_W^0 r)}{M_W r} + 
\frac{g^4}{16\pi^2} 
\left(A+B\log(\mu^2/M_W^2)\right),
\end{eqnarray}
where $M_W^0 =M_W-\delta M_W$, with $\delta M_W$ the one-loop mass correction. 
Since $\delta M_W$ is scale dependent, so is $M_W^0$.
A and B are functions of the distance $r$ and $R_{HW}=M_H/M_W$. 
We choose $M_W=80$GeV.
Fig.\ 2 shows the  dependence of $A$ and $B$ on the dimensionless distance 
$rM_W$ for $R_{HW}=0.8314$ (corresponding to the end point of the first 
order finite  temperature phase transition \cite{Fod8}), while Fig.~3 
shows the $R_{HW}$ dependence for $r=M_W^{-1}$.

\begin{figure}
\begin{center}
\epsfig{file=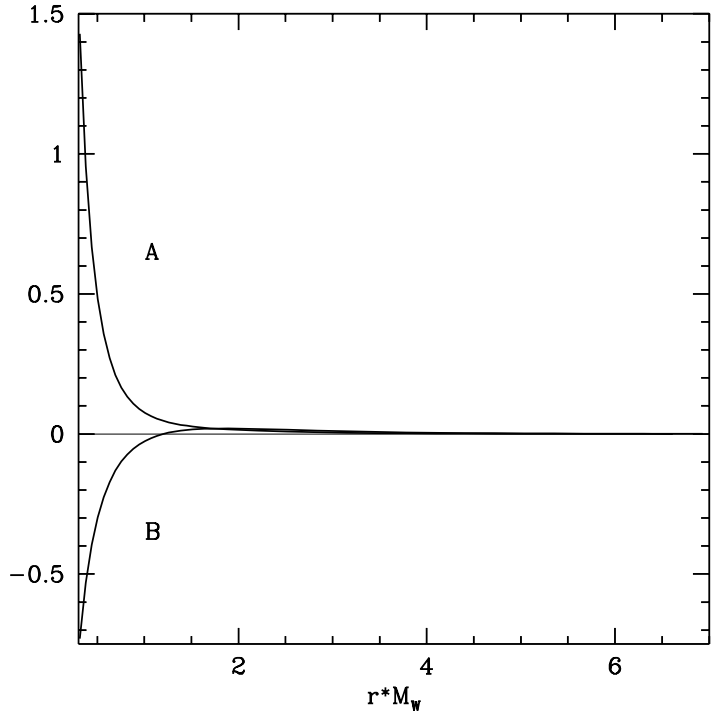,width=10.0cm}
\caption{\label{pot_fig}
{The coefficients of $g^4/(16 \pi^2)$---curve A---and of $g^4/(16\pi^2) 
\log(\mu^2/M_W^2)$---curve B---of the static potencial Eq.\ (\ref{pot}) as a function of 
distance times W mass. $R_{HW}$=0.8314.
}}
\end{center}\end{figure}

\begin{figure}
\begin{center}
\epsfig{file=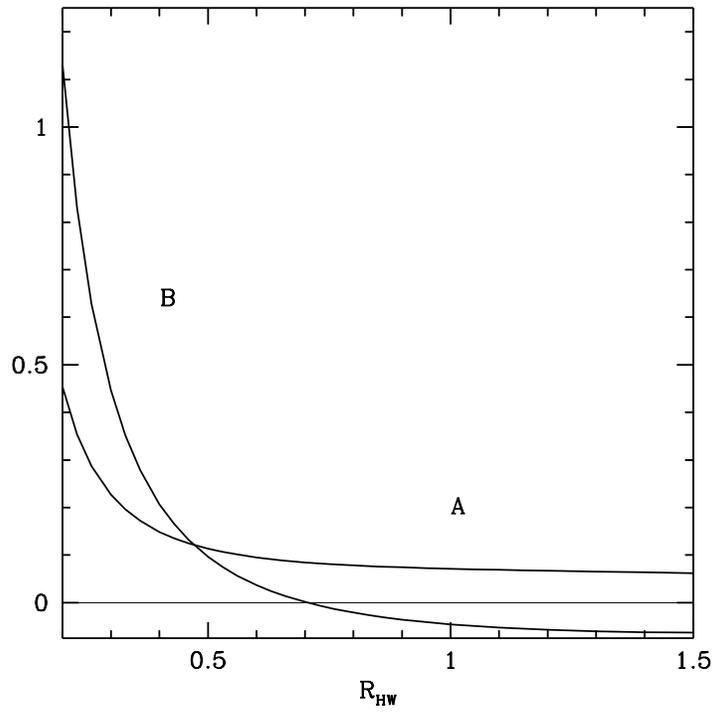,width=10.cm}
\caption{\label{pot_R_fig}
{The coefficients of $g^4/(16\pi^2)$---curve A---and of $g^4/(16\pi^2) 
\log(\mu^2/M_W^2)$---curve B---of the static potencial Eq.\ (\ref{pot}) 
as a function of $R_{HW}$. 
The distance is $M_W^{-1}$.
}}
\end{center}\end{figure}

\section{Relation of the continuum version of the lattice coupling constant
definition to the ${\mathbf{\overline {\mathbf {MS}}}}$ coupling constant }

Since we wish to compare results of lattice simulations and continuum perturbation theory 
calculations, it is an essential point to define the SU(2) gauge coupling in the 
same way in both cases. However, in continuum perturbation theory the 
$\overline {\rm {MS}}$ running coupling constant at a given renormalization scale 
is more natural (as used in Eqs.\ (\ref{mom_pot},\ref{pot}), too), while in 
lattice simulations other definitions are applied. 
Therefore we have to establish the relation between the coupling constants.

The lattice definition of the coupling constant is given in \cite{Fod1}. 
Note that we are using the local version below. For the reader's convenience we 
recall this definition (inspired by \cite{Som}).\\
First rectangular Wilson loops of size (r,t) are measured. 
Extrapolating to large $t$ and dividing the logarithm by $-t$ one gets
the static potential in the $t \rightarrow \infty$ limit as a function of r.
The nonperturbative lattice static potential is fitted by a finite lattice 
version of the Yukawa potential with four parameters (for details cf.\ 
\cite{Fod1}). One of these parameters is the mass in the exponential of the 
Yukawa potential, which is usually called the screening mass. The gauge 
coupling  at distance $r$ is defined as the ratio of the discrete 
$r$ derivative of the lattice simulated nonperturbative potential 
and the discrete derivative of the tree-level 
lattice Yukawa potential normalized by the square of the tree-level coupling 
and with the mass parameter $M_{\mathrm{lattice}}$ identified with the screening mass.
In practice $g^2_{\mathrm{lattice}} (M_{\mathrm{lattice}}^{-1} )$ is determined
and is called the local renormalized gauge coupling constant on the lattice.
The lattice results at various Higgs masses are collected in Table 1.
Data are from \cite{4d-asym},\cite{Fod8},\cite{Fod1}, and \cite{Fod3}.

To follow  the above procedure in the case of the continuum perturbative 
determination of the renormalized gauge coupling, we performed a fit of the 
one-loop potential with a tree-level Yukawa potential plus a constant term. 
The parameters of the fit are the coupling constant, the mass in the 
exponent (perturbative ``screening mass'' $M_{\rm screen}$) and the constant. 
For the various values of Higgs mass we performed the fit in the same $r$ 
range as used in the lattice studies and took the errors of the fitted function
to be proportional to the errors of the potential obtained in lattice simulations. 
$g_R^2 (r )$
is then determined by taking the ratio of the derivatives with respect to 
$r$ of the one-loop potential and the tree-level potential normalized by the 
square of the tree-level coupling, i.e.\ we have
\begin{equation}\label{g2_def}
g_{R}^2 (r)=\frac{1}{C_F}\frac{\dst\frac{d}{dr}\left[-V(r)\right]}
{\dst\frac{d}{dr} \int \frac{d^3 k}{(2\pi)^3}\frac{\exp (i{\vec k}\cdot {\vec r})}
{k^2 + M_{\rm screen}^2 }},
\end{equation}
with $V(r)$ given by Eq.\ (\ref{pot}), $C_F=3/4$, and $M_{\rm screen}$  obtained 
from the fit. 
Since $M_{\rm screen}-M_W^0=O(g^2)$, for distances satisfying 
$M_{\rm screen} -1/r =O(g^2) $ we 
can put Eq.\ (\ref{g2_def}) into the form
\begin{eqnarray}\label{g2_res}
g_{R}^2 (r)= g_{\overline {\rm {MS}}}^2 (\mu)\left(1+\frac{1}{2}
\left(1-\frac{M_W^0}{M_{\rm screen}}\right)\right)+\frac{g_{\overline {\rm {MS}}}^4 (\mu)}
{16\pi^2} \left(C+D \log \frac{\mu^2}{M_W^2} \right).
\end{eqnarray}
$C$ and $D$ are functions of $R_{HW}$ and $M_{\rm screen}$, their 
values are tabulated in Table 2 for $M_{\rm screen}=M_W=80$GeV.

\begin{table}[htb]
\begin{center}
\begin{tabular}{|c|c|c|c|c|}
\hline
 $R_{HW}$ & .2049 & .4220 & .595  & .8314 \\
 \hline
 $T_c$  (GeV)  & 38.3 & 72.6 & 100.0 & 128.4\\
 \hline
 $M_{\mathrm{lattice}}$ (GeV)& 84.3(12) & 78.6(2) & 80.0(4) & 76.7(24) \\
 \hline
 $g^2_{\mathrm{lattice}} (M^{-1} )$ & .5630(60) & .5788(16) & .5782(25) & .569(4) \\
 \hline
 $M_{\rm screen}$ (GeV) &74.97 & 80.44 & 80.70 & 81.77 \\
 \hline
 $g^2_{\overline {\rm {MS}}} (T_c )$ &0.540 & 0.592 & 0.585 & 0.570 \\
 \hline
 $g_{\overline {\rm {MS}}}^{2,Laine} (T_c)$ & 0.589 & 0.589 & 0.579 & 0.562
 \\
 \hline
 \end{tabular}
 \caption{\label{couplings}
 Various quantities calculated for values of $R_{HW}$ used in lattice simulations.
 For more explanation see the text. As usual the numbers in the parentheses 
 denote the errors in units of the last decimals. The errors of the 
 different gauge couplings
 are dominated by the lattice simulation errors (fourth row), therefore
 we did not indicate them in rows 6 and 7.
 }
 \end{center}
 \end{table}

In this procedure we have to choose the  gauge coupling in the one-loop potential
so that $g_R^2 (M_{\rm screen}^{-1} )$ reproduces the lattice result (third row of 
Table 1) for the appropriate value of the Higgs mass. 
For our applications (thermodynamical quantities at and around the critical 
temperature $T_c$ of the first order electroweak phase transition) the scale
of the one-loop potential is chosen to be $T_c\approx 2M_H$, where 
$M_H$ is the  Higgs boson mass at zero temperature.  
 Thus the  
gauge coupling appearing in the one-loop potential is actually the 
${\overline {\rm {MS}}}$ gauge coupling 
at scale $T_c$. The ${\overline {\rm {MS}}}$ gauge coupling values obtained from this 
procedure are given in the sixth row of Table 1.

Another definition of the continuum perturbation theory one-loop 
``renormalized gauge coupling'' at distance $r$ is given in \cite{Laine}. 
It reads
\begin{eqnarray}\label{Laine_def}
g_{R,Laine}^2 (r)=\frac{1}{C_F}\frac{\dst{\frac{d}{dr}\left[-V(r)\right]}}
{\dst{\frac{d}{dr} \int \frac{d^3 k}{(2\pi)^3}\frac{\exp (i{\vec k}\cdot {\vec r})}
{k^2 + M^2 }}},
\end{eqnarray}
where $M$ is a free mass parameter satisfying $M-M_W \propto g^2$. For this $M$ it
is possible to show that $g_{R,Laine}^2 (M^{-1})$ can be expressed in terms of 
$g_{\overline {\rm {MS}}}^2 (M_W)$,  
(where $M_W$ is the physical (one-loop) pole mass) and all the scale 
dependence is included in $g_{\overline {\rm {MS}}}^2 (M_W)$. Assuming 
$M=M_W$, the numerical difference between this definition and ours is
small. However, we believe that it is our definition which is the closest
conceivable to the local renormalized lattice gauge coupling of 
\cite{Fod1}. In Table 1 (last row) we give 
$g_{\overline {\rm {MS}}}^{2,Laine} (T_c)$
as calculated using Eq.\ (\ref{Laine_def}), equating $g_{R,Laine}^2 (M_W^{-1})$ 
with the values of the lattice simulation 
results  $g_{\mathrm{lattice}}^2 (M_{\mathrm{lattice}}^{-1})$ and using the 
renormalization group equation to extrapolate to the scale $T_c$.

\begin{table}[htb]
\begin{center}
\begin{tabular}{|c|c|c|}
\hline
 $R_{HW}$ & $C$ & $D$ \\
 \hline
 0.2 &-41.54 & -22.19\\
 \hline
 0.3 &-8.26 & -6.58\\
 \hline
 0.4 &-6.47 & -1.12\\
 \hline
 0.5 & -5.66& 1.39\\
 \hline
 0.6 & -5.23& 2.74\\
 \hline
 0.7 &-4.98 & 3.55\\
 \hline
 0.8 & -4.83& 4.06\\
 \hline
 0.9 &-4.72 & 4.39\\
 \hline
 1.0 & -4.65& 4.62\\
 \hline
 1.1 & -4.59& 4.78\\
 \hline
 1.2 & -4.54& 4.89\\
 \hline
 1.3 & -4.50& 4.98\\
 \hline
 1.4 & -4.45& 4.98\\
 \hline
 1.5 &-4.40 & 5.01\\
 \hline
 \end{tabular}
 \caption{\label{coupling_res}
 The coefficients $C$ and $D$ defined in
 Eq.\ (\ref{g2_res}) as a function of $R_{HW}$.
 }
 \end{center}
 \end{table}

\section{Comparison of physical observables determined in lattice simulations 
with perturbative predictions}

In the previous section we presented a
calculation connecting the renormalized gauge coupling constant of the
${\overline {\rm {MS}}}$ scheme and $g_R^2$ obtained from the static potential
at different distances. In this section we compare the lattice results
and the perturbative predictions for the finite temperature electroweak
phase transition.  Lattice Monte Carlo simulations provide a 
well-defined and systematic approach to study the features of the finite
temperature electroweak phase transition.
During the last years large scale numerical simulations have been carried out
in four dimensions in order to clarify non-perturbative details
\cite{4d-asym},\cite{Fod8},\cite{Fod1},\cite{Fod3}.    
Thermodynamical quantities (e.g.\ critical
temperature, jump of the order parameter, interface tension, latent heat)
have been determined and extrapolation to the continuum limit has been performed in
several cases. Nevertheless, it has proven difficult
to compare the perturbative and the lattice results, because the
perturbative approach used the ${\overline {\rm {MS}}}$ scheme for the gauge coupling,
whereas the lattice determination of the gauge coupling has been based on
the static potential. The main reason for performing the one-loop calculation
of the static potential is this kind of comparison.

In this paper we use the published perturbative two-loop result for the
finite temperature effective potential of the SU(2)-Higgs model
\cite{FH94}. Note that the numerical evaluation of
the one-loop temperature integrals gives a result which agrees with the
approximation based on high temperature expansion within a few percent.
The reason for this is that the perturbative expansion
up to order $g^4,\lambda^2$ corresponds to a high temperature
expansion, which is quite precise for the Higgs boson masses we studied.
It is known that the perturbative loop expansion becomes unreliable
for Higgs masses above approximately 50 GeV (e.g.\ resummed perturbation 
theory fails
to predict the end-point of the electroweak phase transition, thus it
gives a first order phase transition for arbitrarily large
Higgs boson masses). In the physically relevant range of the parameter space the
electroweak phase transition can only be understood by means of
non-perturbative methods. Therefore it is particularly
instructive to see quantitatively
how the perturbative and the lattice results agree for
small Higgs boson masses and how they differ for larger ones.

Since the finite temperature electroweak phase transition is
fairly strong for Higgs boson masses below 50 GeV, 
lattices with symmetric lattice spacings were used for $M_H
\approx 16$ GeV, $M_H \approx 34$ GeV  and $M_H\approx 48$ GeV.
The phase transition gets weaker for larger Higgs boson masses, therefore
Monte Carlo simulations for masses near the W-boson mass
are technically difficult. For this parameter region different
lattice spacings were used in the temporal and the spatial
directions. For this type of lattice regularization the
approach to the continuum limit is somewhat slower; however,
even in this case it was possible to perform
a continuum limit extrapolation for $M_H\approx 67$ GeV.

In lattice simulations the gauge coupling constant are determined from
the static potential, whereas masses are extracted from correlation functions.
On the one hand the calculation of the previous section connects the gauge
coupling definitions between the $\overline{\rm {MS}}$ scheme and the
scheme based on the static potential. On the other hand one can
use the zero temperature effective potential in order to
include the most important mass renormalization effects. The Higgs boson
mass obtained from the asymptotics of the correlation function corresponds 
to the physical mass determined by the pole of the propagators, i.e.\ the 
solution of $p^2-M^2=\Pi (p^2)$, where $\Pi (p^2)$ is the self-energy. The
effective potential approach
suggested by Arnold and Espinosa \cite{AE93} approximates $\Pi(p^2)$ by
$\Pi (0)$ in the above dispersion relation. It has been argued that the
difference between the two expressions is of order $g^5 v^2$  ($v$ is the
zero-temperature vacuum expectation value), which
does not affect our discussion. In this scheme the correction to the
$\overline{\rm {MS}}$ potential reads
\begin{equation}
\delta V={\varphi^2 \over 2} \left( \delta m^2+ {1 \over 2\beta^2}
\delta\lambda\right) + {\delta\lambda \over 4} \varphi^4,
\end{equation}
where
\begin{equation}
\delta m^2 = {9g^4v^2 \over 256 \pi^2},\ \ \ \ \ \
\delta\lambda=-{9g^4\over 256\pi^2}\left(\log\frac{M_W^2}{\mu}+{2\over
3} \right).
\end{equation}
Here ${\mu}$ is the renormalization scale and $M_W$ is the W-boson
mass at $T=0$. The above
notation corresponds to a tree-level potential of the form
$m^2 \varphi^2/2+\lambda \varphi^4/4$. Note that this treatment is
analogous to previous comparisons of the perturbative and lattice results
\cite{BFH95}.

In \cite{4d-asym},\cite{Fod8},\cite{Fod1},\cite{Fod3}  several observables 
were determined, including renormalized masses at zero temperature ($M_H$,
$M_W$), critical temperatures ($T_c$), jumps of the order parameter 
($\varphi_+$), latent heats ($Q$) and surface tensions ($\sigma$) for
different Higgs boson masses. As usual, the dimensionful quantities were
normalized by the proper power of the critical temperature. (This convention is 
adapted in the present paper, too.) The simulations
were performed on $L_t=2,3,4,5$  lattices ($L_t$ is the temporal
extension of the finite-temperature lattice) and whenever
it was possible a systematic
continuum limit extrapolation was carried out assuming standard $1/a^2$
corrections for the bosonic theory.

%\renewcommand{\arraystretch}{1.5}\label{comp}
%\bigskip \\
\begin{table}  \begin{center}
\label{comp}
\begin{tabular}{|c|}
\hline
 $M_H$\\ \hline $g_R^2$  \\
 \hline
  \begin{tabular}{c|c}
  $T_c/M_H$ \hspace{-5.5pt} &
   \begin{tabular}{c}
   pert \\
   \hline
   nonpert
  \end{tabular}
 \end{tabular} \\
 \hline
  \begin{tabular}{c|c}
 $\varphi_+/T_c$ \hspace{-1.5pt} &
  \begin{tabular}{c}
   pert \\
   \hline
   nonpert
  \end{tabular}
 \end{tabular} \\
 \hline
  \begin{tabular}{c|c}
  $Q/T_c^4$ \hspace{1.5pt} &
   \begin{tabular}{c}
   pert \\
   \hline
   nonpert
   \end{tabular}
  \end{tabular} \\
\hline
\begin{tabular}{c|c}
$\sigma/T_c^3$ \hspace{3.6pt} &
\begin{tabular}{c}
pert \\
\hline
nonpert
\end{tabular}
\end{tabular}\\
\hline
\end{tabular}
\begin{tabular}{|c|c|c|c|}
\hline
16.4(7)          & 33.7(10)         & 47.6(16)         & 66.5(14) \\
\hline
0.561(6)         & 0.585(9)         & 0.585(7)         & 0.582(7) \\
\hline
2.72(3)          & 2.28(1)          & 2.15(2)          & 1.99(2)  \\
\hline
2.34(5)          & 2.15(4)          & 2.10(5)          & 1.93(7)  \\
\hline
4.30(23)         & 1.58(7)          & 0.97(4)          & 0.65(2)  \\
\hline
4.53(26)         & 1.65(14)         & 1.00(6)          & 0 \\
\hline
0.97(7)          & 0.22(2)          & 0.092(6)         & 0.045(2) \\
\hline
1.57(37)         & 0.24(3)$^*$      & 0.12(2)          & 0 \\
\hline
0.70(10)         & 0.067(6)         & 0.022(2)         & 0.0096(5) \\
\hline
0.77(11)         & 0.053(5)$^*$    & 0.008(2)$^*$     & 0 \\
\hline
\end{tabular}
\end{center} 
\caption[]{ Comparison of the perturbative and the lattice results.
The Monte Carlo data are from
\cite{4d-asym},\cite{Fod8},\cite{Fod1},\cite{Fod3} 
 (in some cases we have refined
the analysis in order to have a more accurate lattice prediction).
Note that for the mass
of the W boson---the dimensionful quantity setting the scale of 
the theory---80 GeV is used.}
\end{table}

The statistical errors of these observables are normally determined by
comparing statistically independent samples. 
Jackknife and bootstrap techniques were used \cite{14} and correlated 
fits were performed \cite{15} to obtain reliable estimates of the statistical 
uncertainties. The systematic errors due to
finite lattice spacing can be obtained by $1/a^2$ extrapolation. In cases where 
it was possible to carry out the continuum limit extrapolation we saw  
that the difference between the $L_t=2$ and
the $L_t=3$ data was a fairly good estimator of the systematic error.
Whenever the data did not make it possible to carry out
the systematic extrapolation the difference between the $L_t=2$ and
the $L_t=3$ results was used to estimate the systematic error. As a
conservative estimate we added the statistical and systematic errors linearly.
For some of the quantities only $L_t=2$ data exist. In these cases
only the statistical errors are listed and an asterisk is used
in Table 3 as an indication. A correct
comparison has to include errors on the parameters used in the perturbative
calculation. These uncertainties are connected with the fact that neither the
Higgs boson mass nor the gauge coupling constant
can be   determined exactly in lattice simulations. Including  these errors, 
the perturbative prediction for an observable is rather an interval than one 
definite value. 

To obtain a better measure of the correspondence between perturbative and 
nonperturbative results, and to incorporate their errors, one introduces ``pulls''
defined by the expression
\begin{equation}
\mathrm{pull} = \frac{{\mathrm{perturbative\ mean}} - {\mathrm{nonperturbative\ mean}}}
{{\mathrm{perturbative\ error}} + {\mathrm{nonperturbative\ error}}}.
\end{equation}
The four different pulls at different Higgs boson masses are tabulated in Table 4 and
plotted in Fig.~4. 
For the sake of convenience, we used the shorthand $P_T=\mathrm{pull\ of\ } T_C/M_H$, 
$P_\phi=\mathrm{pull\ of\ } 
\varphi_+/T_C$, $P_Q=\mathrm{pull\ of\ } Q/T_C^4$, and $P_\sigma=\mathrm{pull\ of\ }
\sigma/T_C^3$.
\begin{table}[htb]
\begin{center}
\begin{tabular}{|c|c|c|c|c|}
\hline
$m_H$ (GeV) & 16.4(7) & 33.7(10) & 47.6(16) & 66.5(14) \\
\hline
$P_T$       & 4.75    & 2.60     & 0.71     & 0.67 \\
\hline
$P_\phi$ & 0.47    & -0.33    & -0.3     & 32.5 \\
\hline
$P_Q$	    & -1.36   & -0.4    & -1.08    & 22.5 \\
\hline
$P_\sigma$  & -0.33   & 1.27	 & 3.5	    & 19.2 \\
\hline
\end{tabular}
\caption{Values of the four different pulls for various Higgs boson masses}
\end{center}
\end{table}
\begin{figure}
\begin{center}
\hspace{-1cm}
\epsfig{file=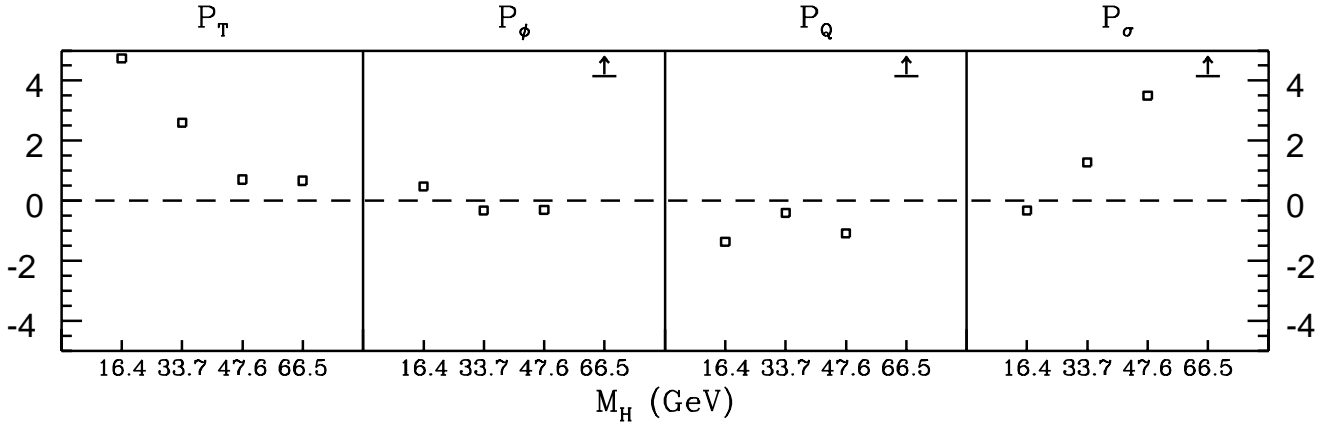,width=18cm}
\caption{\label{fig4}
``Pulls'' plotted against the Higgs mass. 
%Ha az az abra kerul be...
Arrows indicate values outside the
interval $[-5,5]$.}
\end{center}
\end{figure}

The quantity which has the smallest pull even for large Higgs boson masses 
is $T_c /M_H $. A quadratic fit was performed to this quantity as a function 
of $R_{HW}$. The result is
\begin{equation}
\frac{T_c}{M_H}=2.494-0.842 R_{HW} + 0.223 R_{HW}^2.
\end{equation}

In \cite{Fod8} the end point result for the four-dimensional SU(2)-Higgs 
model was perturbatively converted to the Standard Model. In this step 
the deviation between the two definitions of the gauge coupling 
($g^2_{\overline {\rm {MS}}}$ and the one based on the static potential) 
was neglected. The estimated uncertainty due to this simplification was  
included into the systematic error of the end point Higgs mass. 
Using the results of the present paper 
we can also refine the value of the Standard Model end point Higgs mass. This
is done by perturbatively taking into account fermions and the U(1) factor 
of the Standard Model (for details see the fourth paper of \cite{3d-sim}). 
The improvement is established by precisely converting  the lattice simulation 
renormalized gauge coupling of the SU(2)-Higgs model 
into the perturbative $g^2_{\overline {\rm {MS}}} (M_W )$. 
The new value of the Standard Model end point Higgs mass is $72.1 \pm 1.4$GeV. 
This does not deviate much from the old value $72.4 \pm 1.7$GeV 
of \cite{Fod8}. However, 
the error is smaller, since the uncertainty arising from the gauge coupling 
definitions is eliminated.

\section{Summary }

In this paper we presented the one-loop static potential in the
SU(2)-Higgs model. This calculation is in agreement with \cite{Laine}.
Using the potential it was possible to connect the gauge coupling
constant used in finite temperature field theory and in lattice
simulations. As expected the numerical difference between the
two conventions is not that large, it is within a few percent.
With this connection we were able to perform a precise comparison
between the predictions of perturbative and lattice approaches.

We reanalyzed the existing lattice data and performed a 
continuum limit extrapolation whenever it was possible. The 
relationship between the two definitions of the gauge coupling
constants turned out to be marginal, as the lattice data
have errors, usually larger then this few percent. The only
quantity which is measured so precisely that the definition
of the gauge coupling constant is essential is the ratio of the
critical temperature to the Higgs boson mass. As it has been 
observed already for $M_H \approx 35$ GeV the perturbative
value of $T_c$ is larger than in lattice simulations. 
This sort of discrepancy disappears for larger
Higgs boson masses. A plausible reason for this fact
is the convergence of the high temperature expansion
used in the perturbative approach. In two-loop
perturbation theory one uses the high temperature expansion 
also up to second order, which might be inaccurate for the smallest Higgs 
mass case with temperatures $\approx 50$ GeV and
Higgs field expectation values $\approx 200$ GeV. Nevertheless, 
the observed differences are on the percent level and they do not affect 
the electroweak phase transition significantly. For small
Higgs boson masses (16 and 34 GeV) we expect
similar differences between lattice and perturbative 
predictions for other quatities (the jump of the order parameter,
the latent heat and the surface tension); however, present lattice
data have too large errors and the differences cannot be seen yet. 

The most dramatic differences appear clearly as we get closer to the end point. 
The perturbative approach gives nonvanishing jump of the 
order parameter, nonvanishing latent heat and interface tension, while 
the lattice results suggest rapid decrease of these quantities 
and no phase transition beyond the end point. 

Using the results of the present paper
we refined the value of the Standard Model end point Higgs mass of \cite{Fod8} 
and obtained $72.1 \pm 1.4$GeV.

\vspace{.5cm}

This work was partially supported by
Hungarian Science Foundation Grants under Contract  
No.\ OTKA-T22929-29803-M28413/FKFP-0128/1997.

\vfill

\end{document}